\documentstyle[psfig]{l-aa}
\begin{document}

\thesaurus{ 08.01.2, 08.02.1, 08.03.3, 08.06.1, 08.12.1 }

\title {Simultaneous H$\alpha$, Na~{\sc i} D$_{1}$, D$_{2}$, 
and He~{\sc i} D$_{3}$
observations of a flare on the RS CVn system UX Ari
\thanks{Based on observations made with the Isaac Newton telescope
operated on the island of La Palma by the 
Royal Greenwich Observatory at
the Spanish Observatorio del Roque de Los Muchachos of the
 Instituto de Astrof\'{\i}sica de Canarias on behalf of 
the Science and Engineering Research Council 
of the United Kingdom and the Netherland 
Organization for Scientific Research.
} }

\author{D.~Montes
\and J.~Sanz-Forcada
\and M.J.~Fern\'{a}ndez-Figueroa
\and R.~Lorente
}

\offprints{ D.~Montes}

\institute{Departamento de Astrof\'{\i}sica, 
Facultad de F\'{\i}sicas,
 Universidad Complutense de Madrid, E-28040 Madrid, Spain
\\  E-mail: dmg@ucmast.fis.ucm.es}

\date{Received ; accepted }

\maketitle
\markboth{D. Montes et al.: 
H$\alpha$, Na~{\sc i} D$_{1}$, D$_{2}$, 
and He~{\sc i} D$_{3}$ observations of a flare on UX Ari
}{ }


\begin{abstract}

We present simultaneous H$\alpha$, Na~{\sc i} D$_{1}$, D$_{2}$, 
and He~{\sc i} D$_{3}$ spectroscopic observations 
on the RS CVn system UX Ari.
We have found a dramatic increase in the excess H$\alpha$ emission 
equivalent width by a factor of 2 in an interval of 1 day that indicates
the beginning of a strong flare in this system.
The presence of the He~{\sc i} D$_{3}$ in emission in coincidence with the 
enhancement of the H$\alpha$ emission confirms the detection of a flare. 
The application of the spectral subtraction technique reveals that the 
core of the Na~{\sc i} D$_{1}$ and D$_{2}$ lines are also filled-in by 
chromospheric emission.


\end{abstract}

\section{Introduction}

The RS CVn system UX Ari (HD 21242) is a non-eclipsing double-lined 
spectroscopic binary with spectral type (G5V/K0IV) and orbital period of
6.43791 days. 
It is one of the few RS CVn systems, along with V711 Tau, II~Peg, and DM UMa,
that shows H$\alpha$ consistently in emission.
The extensive Ca~{\sc ii} H \& K and H$\alpha$ observations of UX Ari
(Carlos \& Popper 1971;  Bopp \& Talcott 1978; Nations \& Ramsey 1986;
Huenemoerder et al. 1989; Frasca \& Catalano 1994;
Raveendran \& Mohin 1995; Montes et al. 1995a,b,~c)
show that the chromospheric emission belongs to 
the cooler component of the system, 
however some authors (Simon \& Linsky 1980; Huenemoerder et al. 1989) 
also found a small contribution from the G5V secondary component.

Simon et al. (1980) observed a UV flare on UX Ari and found that 
chromospheric lines were enhanced 2.5 times their quiescent level,
while transition region lines displayed an enhancement by a factor 
of up to 5.5.
They proposed an activity scenario for long-lived RS CVn flares in 
which the component stars  have  large  corotating  flux  tubes  which 
occasionally interact. The resulting magnetic  reconnection  leads 
to flares. 
Recently, Henry \& Newsom (1996) have detected, 
with photometric observations,
an optical flare in this system.

In this paper, we report the detection of an optical flare in UX Ari 
through simultaneous observations of the chromospheric 
H$\alpha$ and He~{\sc i} D$_{3}$ lines. It is the first time 
that the He{\sc i} D$_{3}$ line is observed in emission in UX Ari,
moreover, a filled-in in the 
Na~{\sc i} D$_{1}$ and D$_{2}$ lines have been found.
In Sect. 2 we give the details of our observations and data reduction
and we describe the method of spectral subtraction to obtain the 
active-chromosphere contribution.
In Sect. 3 we describe the behaviour of the chromospheric activity indicators 
and Sect. 4 gives the conclusions.

\begin{figure*}
\psscalefirst
\vspace{-0.5cm}
\hspace{-1.5cm}
{\psfig{figure=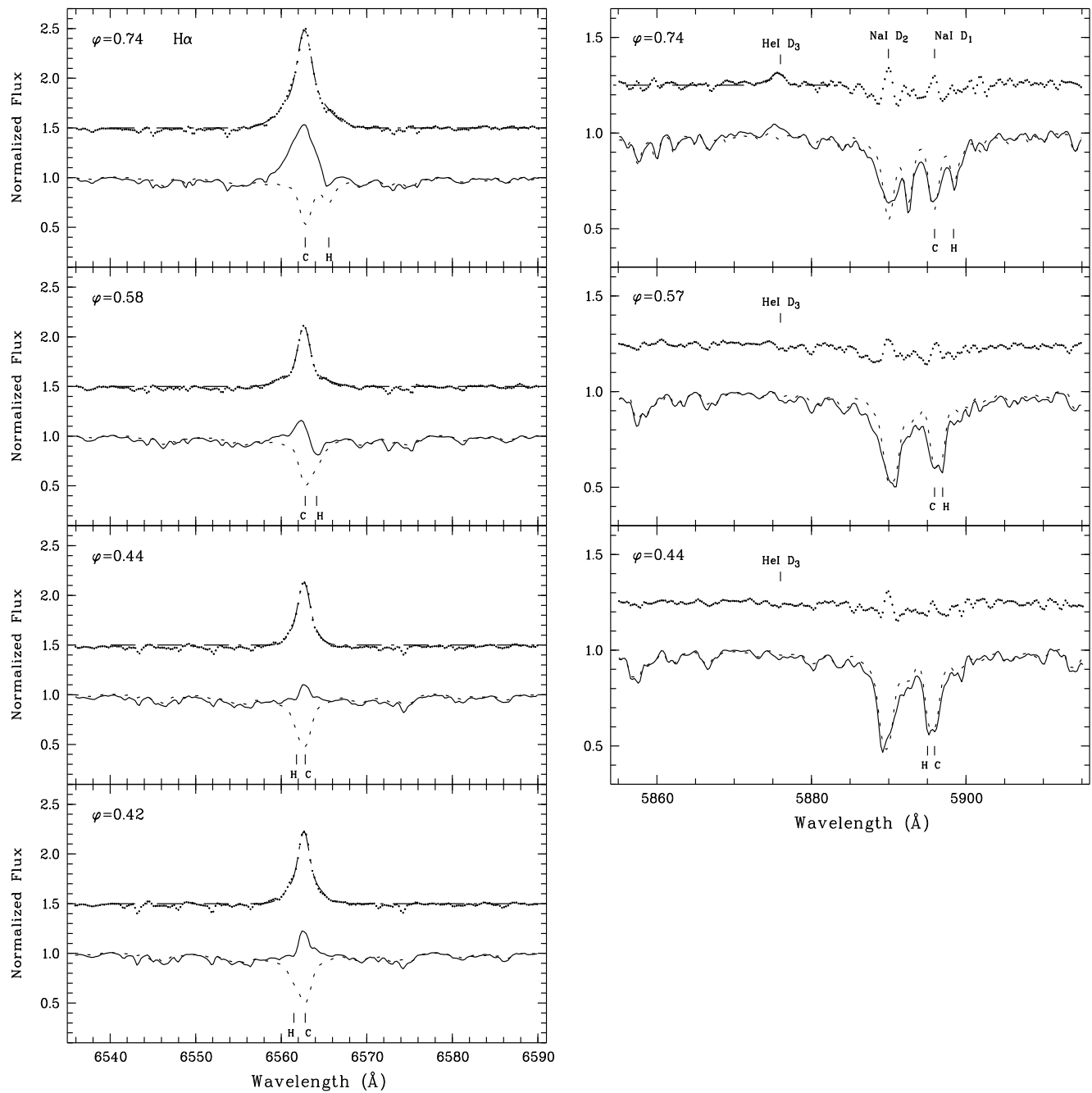,height=27.2cm,width=17.6cm,rheight=18.25cm,rwidth=6cm,angle=270}}
\caption[ ]{H$\alpha$, Na~{\sc i} D$_{1}$, D$_{2}$, 
and He~{\sc i} D$_{3}$ spectra of UX Ari
For each system we plot the observed spectrum (solid-line),
 the synthesized spectrum (dashed-line),
 the subtracted spectrum, additively offset to better display
(dotted line),
 and the two components fit to the subtracted spectrum
(dotted-dashed line).
 \label{fig:plot_paphahe} }
\end{figure*}

\section{Observations and Data Reduction}

Observations of UX Ari in the H$\alpha$ and He~{\sc i} D$_{3}$ line regions
have been obtained during 
three nights (1995 September 13-15) with
the Isaac Newton Telescope (INT) at the
Observatorio del Roque de Los Muchachos (La Palma, Spain)  using the
Intermediate Dispersion Spectrograph (IDS) with grating H1800V, camera 500 
and a 1024~x~1024 pixel TEK3 CCD as detector.
The reciprocal  dispersion  achieved  is
 0.24~\AA/pixel 
which yields a spectral resolution of 0.48~\AA$\ $ and a useful wavelength 
range of 250~\AA$\ $ centered at 6563\AA$\ $ (H$\alpha$) 
and 5876\AA$\ $ (He{\sc i} D$_{3}$) respectively.

The spectra have been extracted using the standard 
reduction procedures in the IRAF
%
%
 package (bias subtraction, 
flat-field division, and optimal
extraction of the spectra). 
The wavelength calibration was obtained by taking
spectra of a Cu-Ar lamp.
Finally, the spectra have been normalized by fitting a continuum.

We have obtained the chromospheric contribution in 
H$\alpha$ and He~{\sc i} D$_{3}$ lines using
the spectral subtraction technique described in
detail by Montes et al. (1995a,~c).
In this method the synthesized spectra were
constructed from artificially rotationally broadened, radial-velocity
shifted, and weighted spectra of inactive stars 
chosen to match the spectral types and luminosity classes 
of both components of the active
system under consideration. 
For UX Ari we have used as reference stars HD20630 (G5V) and HD23249 (K0IV)
and we have adopted the vsin{\it i} (6/37) from Strassmeier et al. (1993), 
(hereafter CABS).

\begin{table*}
\caption[]{H$\alpha$ line measures in the subtracted spectra}
\begin{flushleft}
\scriptsize
\begin{tabular}{ccccccccccccccc}
\noalign{\smallskip}
\hline
\noalign{\smallskip}
\noalign{\smallskip}
        &    &             &     &      &     &                          &  &
\multicolumn{3}{c}{H$\alpha$ broad component} &\ &
\multicolumn{3}{c}{H$\alpha$ narrow component} \\
\cline{9-11}\cline{13-15}                
\noalign{\smallskip}
 {Date} & UT & {$\varphi$} & {I} & FWHM & EW  & {$\log {\rm F}_{\rm S}$} &
L (10$^{30}$) & {I} & FWHM & EW  & & {I} & FWHM & EW \\
        &    &             &     & {\scriptsize (\AA) } & {\scriptsize (\AA)} 
& {\tiny erg cm$^{-2}$ s$^{-1}$} & {\tiny erg s$^{-1}$} &  
 & {\scriptsize (\AA) } & {\scriptsize (\AA)} & & &
{\scriptsize (\AA) } & {\scriptsize (\AA)} \\
\noalign{\smallskip}
\hline
\noalign{\smallskip}
%
%
1995/09/13 & 02:08 & 0.42 & 0.722 & 1.89 & 1.798 & 6.85 & 9.52 & 
0.302 & 3.720 & 1.194 & & 0.431 & 1.317 & 0.605 \\
1995/09/13 & 05:08 & 0.44 & 0.626 & 1.82 & 1.405 & 6.74 & 7.39 & 
0.297 & 2.917 & 0.922 & & 0.344 & 1.316 & 0.483 \\
1995/09/14 & 02:24 & 0.58 & 0.609 & 1.83 & 1.476 & 6.76 & 7.74 & 
0.145 & 5.153 & 0.778 & & 0.468 & 1.512 & 0.753 \\
1995/09/15 & 03:09 & 0.74 & 0.996 & 2.51 & 3.286 & 7.11 & 17.3 & 
0.405 & 5.414 & 2.229 & & 0.580 & 1.723 & 1.069 \\ 
\noalign{\smallskip}
\hline
\noalign{\smallskip}
\end{tabular}
\end{flushleft}
\end{table*}

\section{Chromospheric activity indicators}

Our spectra in the H$\alpha$ line region at orbital 
phases 0.42, 0.44 (September 13), 0.58 (September 14) 
 and 0.74 (September 15) are displayed in Fig.~\ref{fig:plot_paphahe}.
Strong emission from the cool component is present in all the phases. 

The observed spectra exhibit photospheric
absorption lines from both
components shifted in agreement with orbital phases, being more intense the
corresponding to the cooler component, however the relative contribution of 
each component to the total spectrum is not according to 
that obtained with the spectral types and radii given in CABS (0.07/0.93).
To obtain a satisfactory fit between the observed and synthesized spectra
is necessary to take a relative contribution of 0.30/0.70 to the hot and cool 
components. This indicates that the radii given for these stars are wrong 
and/or that when a large fraction of the K0IV star's surface is covered 
by spots, the 
contribution of the G5V star is larger (see Raveendran \& Mohin 1995). 
In the latter case the relative 
contribution of the components may change with time, which is in accordance 
with the different contributions that we used in previous observations 
of this system (Montes et al. 1995b,~c). 


In Table~1 we list the parameters measured in the  
subtracted H$\alpha$ spectra:
the peak emission intensity, I; 
the full width at half maximum, FWHM; 
the excess emission equivalent width, EW; 
the surface flux, $\log {\rm F}_{\rm S}$, 
obtained with the calibration of Pasquini \& Pallavicini (1991)
as a function of (V-R) and  
the luminosity, L, obtained with the radius given in CABS. 

\begin{figure}
\psscalefirst
\vspace{-1.0cm}
\hspace{-0.5cm}
{\psfig{figure=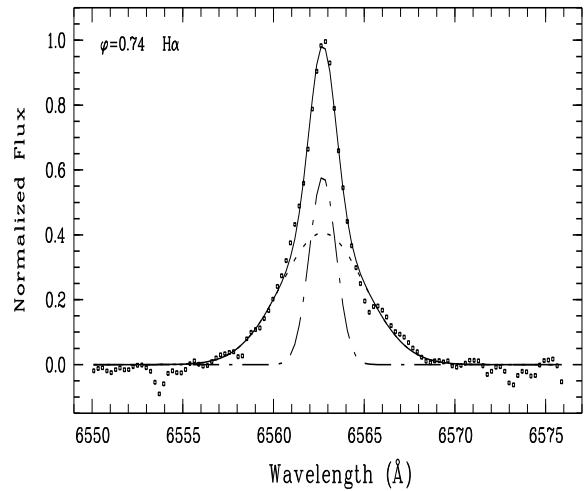,height=35cm,width=15.3cm,rheight=7.5cm,rwidth=2cm,angle=270}}
\caption[ ]{Subtracted H$\alpha$ profile at 0.74 orbital phase (dotted-line).
We have superposed the two Gaussian components fit (solid-line).
The dashed-line represents the broad component and the dotted-dashed-line 
the narrow one.  
 \label{fig:plot_papha_n_b} }
\end{figure}


At 0.74 orbital phase the emission intensity shows a remarkable enhancement
with respect to the other phases.
The excess H$\alpha$ emission equivalent width increases in about a factor 
of 2 
in an interval of 1 day. 
This fact suggests that we have detected a flare since enhancements 
of this amount during flares are typical of chromospheric emission lines
(Simon et al. 1980, Catalano \& Frasca 1994).
The EW obtained in the three other spectra (that we have considered as quiescent 
level) are similar to that measured by us on 1992 December 16 
(Montes et al. 1995b) and similar to the values given in the literature.
The energy released in the flare, 1.7 10$^{31}$ erg~s$^{-1}$ (EW in the flare 
spectrum converted into luminosity) is of the same order of magnitude than the 
maximum H$\alpha$ luminosity observed in the H$\alpha$ flares of other RS CVn 
systems, V711 Tau (Foing et al. 1993) and HK Lac (Catalano \& Frasca 1994).

The subtracted spectra show than the emission corresponds only to the cooler 
component (K0IV) and the H$\alpha$ profile presents broad wings, which are 
much more remarkable in the flare spectrum. 
A Gaussian fit to the subtracted spectrum is not well matched, therefore
we fit the profile using two Gaussian components: a narrow one having a 
FWHM of 60-79 km~s$^{-1}$ and a broad component 
whose FWHM ranges from 133 to 247 km~s$^{-1}$.
In Fig.~ \ref{fig:plot_papha_n_b} we have represented 
the subtracted spectrum at phase 0.74 and the corresponding 
narrow and broad components used to perform the fit.
In Table~1 we give the parameters of the narrow 
and broad Gaussian components used in each spectrum.
As we can see the large changes in the strength of the H$\alpha$ emission,
from the quiescent level to the flare level, occur primarily in the 
broad component, a similar behaviour has been observed in the case of DM UMa 
by Hatzes (1995).
Since in a solar flare large scale mass motions occur, a large flare in UX Ari
may explain the increase of the H$\alpha$ emission and the broad wings 
observed in our spectrum.

In the Na~{\sc i} D$_{1}$, D$_{2}$, and He~{\sc i} D$_{3}$ lines region 
we have obtained three spectra at orbital 
phases 0.44 (September 13), 0.57 (Sept. 14) and 0.74 (Sept. 15), being 
quasi-simultaneous to the H$\alpha$ observations.
The He~{\sc i} D$_{3}$ line appears in emission only at 0.74 phase 
(see Fig.~\ref{fig:plot_paphahe})
corresponding to the increase of the emission observed in the H$\alpha$ line. 
This fact supports the detection of a flare-like event, since in the Sun
the He~{\sc i} D$_{3}$ line appears as absorption in plage and weak flares 
and as emission in strong flares (Zirin 1988). 
The He~{\sc i} D$_{3}$ line has been also observed in emission 
at 0.22, 0.26, 0.77 orbital phases in the RS CVn system II~Peg 
 by Huenemoerder \& Ramsey (1987) and  Huenemoerder et al. (1990).
We have also seen in II~Peg the He~{\sc i} D$_{3}$ line in emission in one 
spectrum at 0.76 orbital phase taken on 1995 September 14 
(paper in preparation).  

We have applied to this spectral region 
the spectral subtraction technique in the same way as for
the H$\alpha$ line. The excess He~{\sc i} D$_{3}$ emission EW obtained in the 
flare spectrum ($\varphi$=0.74) is 0.140 \AA.
In the other two spectra only a very small absorption seems to be present. 
In the case of the Na~{\sc i} D$_{1}$ and D$_{2}$ lines we observe a filled-in 
absorption lines from the cool component in the three spectra, which is 
slightly larger in the flare spectrum (see Fig.~1).
These lines have been observed in emission or as a filled-in 
in red dwarf flare stars (Pettersen et al. 1984; Pettersen 1989). 
As Na~{\sc i} resonance lines are collisionally-controlled in the 
atmospheres of late-type stars, the filled~-~in observed may 
provide information about chromospheric emission.

\section{Conclusions}

We have shown here that the UX Ari binary system exhibits a strong increase
in the H$\alpha$ EW at 0.74 orbital phase. 
In the same phase the He~{\sc i} D$_{3}$ appears in emission, and the
Na~{\sc i} D$_{1}$ and  D$_{2}$ lines show a larger filled-in than in the other 
phases. This behaviour yields us to the conclusion that we have detected a 
flare which released about 1.7~10$^{31}$~erg~s$^{-1}$ in the H$\alpha$ line.

According to the magnetic loop model for UX Ari 
described by Simon et al. (1980) in which the component stars  
have  large  corotating  flux  tubes  which 
occasionally interact, the resulting magnetic  reconnection  leads 
to flares. 
When the emitting regions are seen in projection against the stellar disk 
the He~{\sc i} D$_{3}$ line appears in absorption, when the regions are 
observed off the stellar limb He~{\sc i} D$_{3}$ can be seen in emission. 
The orbital phase (0.74) of our UX Ari observation and its symmetrical position 
are the most favourable to observe the He~{\sc i} D$_{3}$ line in emission.
The other He~{\sc i} D$_{3}$ observations in emission in II~Peg are also seen 
at these orbital phases. 

This fact together with the strong increase
in the H$\alpha$ EW and the larger filled-in of the 
Na{\sc i} D$_{1}$ and D$_{2}$ lines at 0.74 orbital phase
support the presence of a 
flare in the UX Ari binary system.

Photometric observations of this system indicate that the light curve 
minimum (due to cool spots) normally occurred between phase 0.6-0.9 
(Raveendran \& Mohin 1995 and references therein), 
on the other hand Rodon\`{o} \& Cutispoto (1992) observed an anticorrelation 
between the V light curve and the U-B and B-V colours variations that is not 
consistent with the expected variations produced by the presence of cool spots 
alone, and might be due to continuous flaring activity  or bright plage-like 
regions associated with the spots.
Our flare observation occurs at phase 0.74, so the flare eruption could be 
produced (as have been observed in the largest solar flares)
by the interaction of new emerging magnetic flux with old magnetic 
structures associated to the spot groups present in the star at this phase.


\begin{acknowledgements}

This work has been supported by the Universidad Complutense de Madrid
and the Spanish Direcci\'{o}n General de Investigaci\'{o}n 
Cient\'{\i}fica y  T\'{e}cnica (DGICYT) under grant PB94-0263.

\end{acknowledgements}




\end{document}